\documentclass[pre,twocolumn]{revtex4}
\usepackage{epsfig}
\usepackage{bm}
\usepackage{amsmath}

\begin{document}

\title{MHD dynamo in swirling turbulence: from deterministic to helical distributed chaos }

\author{A. Bershadskii}

\affiliation{
ICAR, P.O. Box 31155, Jerusalem 91000, Israel
}

\begin{abstract}

 Using results of laboratory experiments, direct numerical simulations, geomagnetic and solar observations, it is shown that high moments of helicity distribution can dominate power spectra of the magnetic field generated by the magnetohydrodynamic (MHD) dynamo in swirling turbulence even for the cases with zero global helicity. The notion of helical distributed chaos has been used for this purpose.

\end{abstract}

\maketitle

\section{Introduction}

   The swirling flows are characterized by strong (local) helicity and differential rotation, which are typical properties of the flows in the stars' and planets' interiors. In the case of electrically conducting fluids, these properties (at certain conditions) can strongly intensify the conversion of the kinetic energy of the fluid's motion into magnetic energy and support the magnetohydrodynamic (MHD) dynamo.\\
   
   It is known that the MHD dynamo is exited due to the nonlinear instabilities and is developed through the deterministic chaos states (see for instance Ref. \cite{yvw} and references therein). For bounded and smooth dynamical systems one of the simplest ways to determine the presence of deterministic chaos is to compute their power spectrum. The exponential frequency spectrum 
$$
E(f) \propto \exp-(f/f_c)    \eqno{(1)}
$$   
is a good indication in this case \cite{oh}-\cite{mm}.\\

\begin{figure} \vspace{-0.65cm}\centering \hspace{-1.1cm}
\epsfig{width=.48\textwidth,file=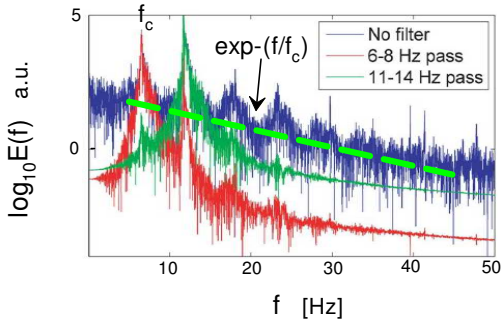} \vspace{-5.3cm}
\caption{Power spectrum of the magnetic field measured by the Hall probe.} 
\end{figure}
   
   A generalization of the deterministic chaos notion for the systems with randomly fluctuating  characteristic frequency  $f_c$ (the distributed chaos) allows consideration of the turbulent MHD dynamo with stretched exponential spectra
$$
E(f) \propto \exp-(f/f_{\beta}) ^{\beta}.   \eqno{(2)}
$$

   A specific form of the stretched exponential spectra with $\beta =1/2$ observed in the direct numerical simulations, laboratory, geomagnetic and solar observations, has been used in the present paper to confirm that the considered MHD processes are dominated by the high moments of helicity distribution
$$
h={\bf v}\cdot {\boldsymbol \omega},   \eqno{(3)}
$$   
 where ${\bf v}$ and ${\boldsymbol \omega} = [\nabla \times {\bf v}]$ are velocity and vorticity fields,  even for the cases with zero global helicity.
   
 \section{Deterministic chaos in MHD}

    In an experiment, described at the site Ref. \cite{kel}, a solid sphere rotates with a constant angular velocity $\Omega_0$ to produce toroidal flow and a hydrofoil propeller with a constant angular velocity $\Omega_i$ (located in the center of the sphere) which pumps along the vertical (rotation) axis to approximate poloidal flow. A weak axial magnetic field $\bf{B_0}$ was imposed on the flow of the electrically conducting fluid (liquid sodium) filling the sphere. The inducted by the fluid motion magnetic field was measured by a Hall probe mounted near the experiment to exclude the imposed magnetic field as much as possible. \\

  Figure 1 shows, in the linear-log scales, power spectra of the signal obtained by the Hall probe at $\Omega_0/2\pi = 5$Hz, $\Omega_i /2\pi= -13$ Hz . The spectral data were taken from the site Ref. \cite{kel}. The dashed straight line is drawn in Fig. 1 to indicate the exponential spectrum Eq. (1) typical for the chaotic systems. It should be noted that the $f_c$ corresponds to the first dominating peak in the spectrum.\\
  
  In paper Ref. \cite{yvw} results of a direct numerical simulation (DNS) of a subcritical transition to MHD dynamo (without an imposed external magnetic field) at the magnetic Prandtl number $Pr_m$ = 0.5 were reported.  \\
  
\begin{figure} \vspace{-1.6cm}\centering \hspace{-1.1cm}
\epsfig{width=.47\textwidth,file=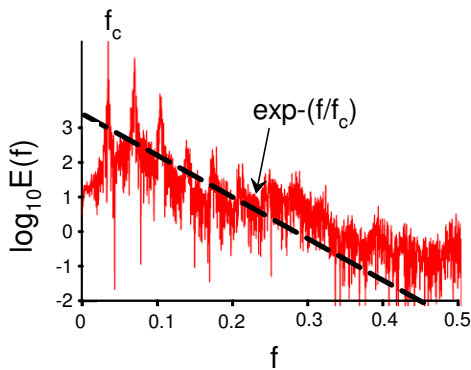} \vspace{-4.65cm}
\caption{Magnetic power spectrum for a chaotic attractor observed in the dynamo-DNS with the Taylor-Green forcing.} 
\end{figure}

\begin{figure} \vspace{-0.5cm}\centering \hspace{-1.3cm}
\epsfig{width=.45\textwidth,file=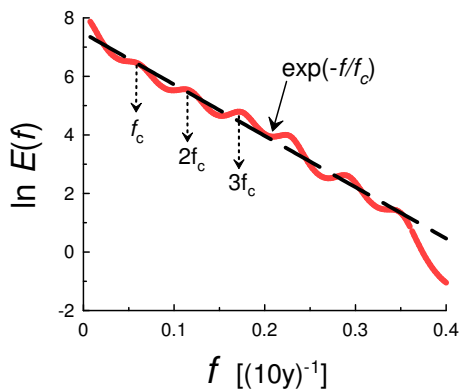} \vspace{-4cm}
\caption{Power spectrum of the magnetic solar activity for the last 11400 years. } 
\end{figure}
  The equations describing dynamics of the incompressible electrically conducting fluid with an associated magnetic field can be written as
$$
 \frac{\partial {\bf v}}{\partial t} = - {\bf v} \cdot \nabla {\bf v} 
    -\frac{1}{\rho} \nabla {\cal P} - [{\bf b} \times (\nabla \times {\bf b})] + \nu \nabla^2  {\bf v} + {\bf f} \eqno{(4)}
$$
$$
\frac{\partial {\bf b}}{\partial t} = \nabla \times ( {\bf v} \times
    {\bf b}) +\eta \nabla^2 {\bf b}  \eqno{(5)} 
$$
$$ 
\nabla \cdot {\bf v}=0, ~~~~~~~~~~~\nabla \cdot {\bf b}=0,  \eqno{(6,7)}
$$
The velocity and normalized magnetic field ${\bf v}$ and ${\bf b} = {\bf B}/\sqrt{\mu_0\rho}$  have the same dimension in the Alfv\'enic units, $ {\bf f}$ is the forcing function. \\

  In this numerical simulation, a mechanical propeller was simulated by the Taylor-Green vortex forcing 
$$
{\bf f}= A
\left[
\begin{array}{c}
\sin(k_0 x)\cos(k_0 y)\cos(k_0 z) \\
-\cos(k_0 x)\sin(k_0 y)\cos(k_0 z) \\
0
\end{array}  
\right]   \eqno{(8)}
$$
in the box geometry without rotation or thermal convection (the forcing wavenumber $k_0 = 2$). The boundary conditions for the Eqs. (4-7) were taken periodic in all three dimensions.\\

   Figure 2 shows a typical magnetic power spectrum obtained in this simulation and corresponding to a chaotic attractor (the spectral data were taken from the Fig. 9c of the Ref. \cite{yvw}). As in Fig. 1, the dashed straight line is drawn in Fig. 2 to indicate the exponential spectrum Eq. (1) typical for the chaotic systems, and the $f_c$ corresponds to the first dominating peak in the spectrum. The authors of the Ref. \cite{yvw} have also noted that the dynamo states observed in their simulation are similar to the transitional dynamo states observed in the VKS dynamo experiment \cite{mon1}. We will return to this experiment with more details below.\\
   
  The modulation and transport of the galactic cosmic rays within the heliosphere is constantly under a strong influence of the Sun’s open magnetic flux, which represents the magnetic solar activity. On the other hand, the $^{14}C$ production rate on the Earth is related to the cosmic ray flux. In paper Ref. \cite{sol} results of a dendrochronologically dated radiocarbon concentrations based reconstruction of the magnetic solar activity for the last 11400 years were reported. \\
  
   Figure 3 shows the power spectrum of the magnetic solar activity for this period. The 10-year averaged data for the spectrum computation were taken from the site Ref. \cite{uso}. The spectrum was computed using the Maximum Entropy Method, specially developed for relatively short data sets. As in Figs. 1 and 2, the dashed straight line is drawn in Fig. 3 to indicate the exponential spectrum Eq. (1) typical for the chaotic systems, and the $f_c$ corresponds to the first dominating peak in the spectrum (see also Ref. \cite{epl}).\\
     
\section{High moments of the helicity distribution}     

   For the case when viscous dissipation can be neglected the dynamics of the mean helicity can be described by equation
$$
\frac{d\langle h \rangle}{dt}  = 2\langle {\boldsymbol \omega}\cdot  (-[{\bf b} \times (\nabla \times {\bf b})] +{\bf f})\rangle \eqno{(9)} 
$$ 
where $\langle...\rangle$ denotes an average over the spatial volume. It is clear that the mean helicity is not an inviscid invariant for this case.  If, however, only the large-scale motions provide the main part to the correlation $ \langle  {\boldsymbol \omega}\cdot  (-[{\bf b} \times (\nabla \times {\bf b})] +{\bf f})\rangle$, the correlation is rapidly decreasing with spatial scales in the chaotic and turbulent flows. As a consequence, the higher moments of the helicity distribution can be considered as inviscid quasi-invariants in this case \cite{lt},\cite{mt}.\\

   To show this, one can divide the spatial domain into a network of the imaginary non-overlapping subdomains  $V_i$ moving with the fluid (in the Lagrangian description) \cite{lt}\cite{mt}. The boundary conditions on the surface of each subdomain are taken in the form ${\boldsymbol \omega} \cdot {\bf n}=0$. Then the moments of order $n$ for the helicity distribution can be then defined as 
 $$
I_n = \lim_{V \rightarrow  \infty} \frac{1}{V} \sum_j H_{j}^n  \eqno{(10)}
$$
where the helicity $H_j$ for the subdomain $V_j$
$$
H_j = \int_{V_j} h({\bf r},t) ~ d{\bf r}.  \eqno{(11)}
$$
   Due to the rapid reduction of the correlation $ \langle  {\boldsymbol \omega}\cdot  (-[{\bf b} \times (\nabla \times {\bf b})] +{\bf f})\rangle$ with spatial scales the subdomains' helicities $H_j$ can be approximately considered as inviscid invariants for the subdomains characterized by the small enough spatial scales. These subdomains should provide the main contribution to the high moments  $I_n$ ($n \gg 1 $) for the turbulent or strongly chaotic flows (cf. \cite{bt}).  Hence, the high moments $I_n$ can be approximately considered as inviscid invariants even when the global helicity $I_1$ cannot. As for the viscous case, the high moments $I_n$ can be considered as {\it adiabatic} quasi-invariants in the inertial range of scales. \\
 
\begin{figure} \vspace{-1.2cm}\centering \hspace{-1.3cm}
\epsfig{width=.45\textwidth,file=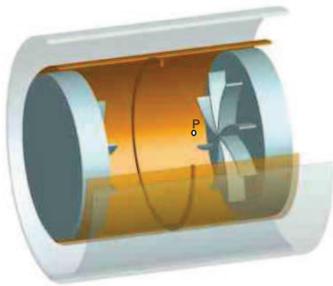} \vspace{-5cm}
\caption{Schematic of the VKS dynamo  experiment configuration. } 
\end{figure}

   It should be also noted that even in the case when the global helicity is equal to zero (due to a spatial symmetry, for instance) the high moments $I_n$ (at least with the even $n$) are non-zero \cite{mt}.  \\ 
   
   The basins of attraction of the chaotic attractors corresponding to the adiabatic invariants $I_n$ are usually different for different values of $n$. Chaotic attractor with a smaller value of $n$ has a thicker basin of attraction than that with a larger value of $n$. Therefore, usually, the flow is dominated by $I_n$ with the smallest value of $n$ for which the $I_n$ can be already considered as a finite adiabatic invariant.\\
 
    In the Alfv\'enic units, $ {\bf b}$ has the same dimension as velocity and, therefore, one can use the dimensional considerations to obtain a relationship between the characteristic values $ {\bf b}_c$ and $f_c$ in the fluid motion dominated by the adiabatic invariant $I_n$
 $$
 b_c \propto  |I_n|^{1/(4n-3)}~ f_c^{\alpha_n}    \eqno{(12)}
$$
  with
 $$
 \alpha_n = \frac{2n-3}{4n-3}   \eqno{(13)}
 $$

  Then for $n \gg 1$ the $\alpha_n \simeq 1/2$.\\

\section{Helical distributed chaos and MHD dynamo}

    For more intense fluid motions (or/and for other boundary conditions) the parameter $f_c$ can have strong fluctuations.  In this case, a more adequate approach should use an ensemble average over the fluctuating parameter
$$
E(f) = \int P(f_c) ~\exp-(f/f_c)~ df_c, \eqno{(14)}
$$  
to compute the power spectrum. 

  The probability distribution $P(f_c)$ can be readily calculated from the Eq. (12) (at $n \gg 1$) if the characteristic magnetic field $b_c$ is normally distributed. Using some simple algebra on can obtain in this case
$$
P(f_c) \propto f_c^{-1/2} \exp-(f_c/4f_{\beta})  \eqno{(15)}
$$
where $f_{\beta}$ is a constant.

     Substituting Eq. (15) into Eq. (14) we obtain
$$
E(f) \propto \exp-(f/f_{\beta})^{1/2}  \eqno{(16)}
$$     
  
  Analogous consideration for the spatial distributed chaos one can find, for instance, in Refs. \cite{b1},\cite{b2}.\\

   In paper Ref. \cite{min} a comparison of the results of the famous von Karman sodium (VKS) experiment on the MHD dynamo in swirling turbulence with a relevant direct numerical simulation have been reported. The von Karman swirling turbulence was produced in a cylindrical vessel (with an inner copper shell and annulus located in the midplane) between two counter-rotating impellers. Figure 4 shows a schematic of the VKS experiment configuration. The magnetic field fluctuations were measured by a Hall probe P located in the bulk of the flow (see Fig. 4).\\
   
   Figure 5 shows in the log-log scales the power spectrum of the self-sustained (MHD dynamo) axial magnetic field fluctuations measured by the probe P as the solid black curve. The same Fig. 5 shows also (as the solid gray curve)  analogous power spectrum for the corresponding signal obtained in a direct numerical simulation made in a spatial box using the Eqs. (4-7) with periodic boundary conditions. The mechanical forcing produced in the VKS experiment by the two counter-rotating impellers (see Fig. 4) was simulated in the DNS by the two Taylor-Green vortices. The frequency axis in the Fig. 5 was normalized by the `forcing' frequency $F_0 = u_{rms}/L$ ($u_{rms}$ is the root mean square of the velocity fluctuations, and 2$L$ is the spatial domain side) for the DNS and $F_0 =10$ Hz (the rotation rate of the impellers) for the VKS experiment. \\
   
   It should be noted that for the VKS experiment and for the corresponding DNS the velocity fluctuations are strong and a well-defined mean velocity is absent in the bulk of the swirling turbulent flow. Therefore, Taylor's `frozen-in' hypothesis cannot be applied to these flows (see, for instance, Ref. \cite{pl}). Hence, the spectra in the Fig. 5 can be interpreted as true temporal ones. The dashed curve in the Fig. 5 indicates the stretched exponential spectrum Eq. (16) corresponding to the helical distributed chaos. \\

\begin{figure} \vspace{-1.1cm}\centering \hspace{-1.3cm}
\epsfig{width=.52\textwidth,file=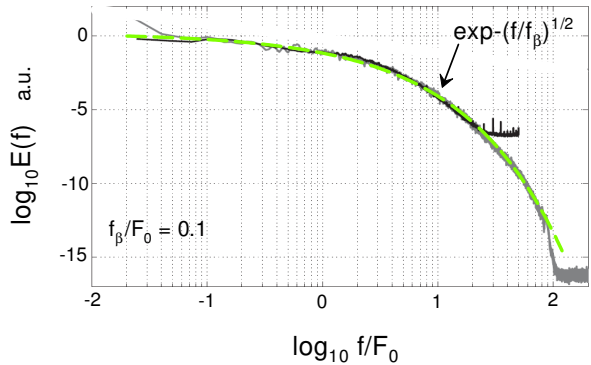} \vspace{-5.8cm}
\caption{Power spectra of the self-sustained (MHD dynamo) axial magnetic field fluctuations measured in the VKS experiment by the probe P (solid black curve), and in the corresponding DNS (solid gray curve).} 
\end{figure}
\begin{figure} \vspace{+2cm}\centering \vspace{-2.5cm}
\epsfig{width=.48\textwidth,file=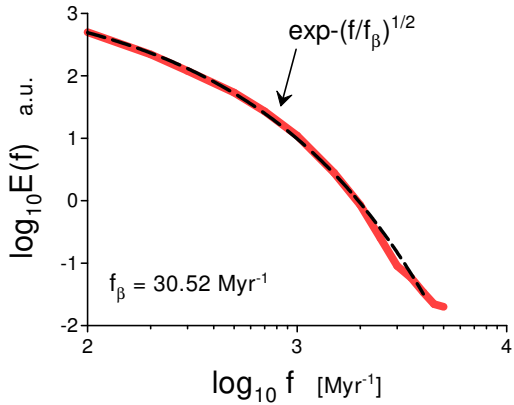} \vspace{-4.6cm}
\caption{A power spectrum of the geomagnetic dipole moment for the time period 0-1 Myr.}
\end{figure}
\begin{figure} \vspace{-1.6cm}\centering
\epsfig{width=.47\textwidth,file=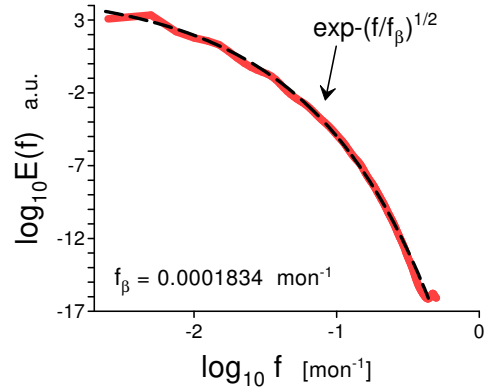} \vspace{-4.1cm}
\caption{ Power spectrum for the reconstructed (cover) time series of the magnetic solar activity for the period 1750-2005 yy.}
\end{figure}

  The Coriolis and buoyancy forces can be included in the term ${\bf f}$ in the Eq. (9) and the consideration of the Sections III and IV can be readily generalized on the rotational and buoyancy-driven fluid motion, i.e. on the realistic geomagnetic and solar dynamos. \\

  The geomagnetic dipole moment (normalized by spatial volume - $V$)
$$
{\boldsymbol \mu} =   \frac{1}{2V} \int [{\bf r} \times {\bf j}]~ dV = \frac{1}{2V} \int [{\bf r} \times (\nabla \times {\bf b})]~ dV  \eqno{(17)}
$$
is usually used to describe the global magnetic field. In the Alfv\'enic units ${\bf b} = {\bf B}/\sqrt{\mu_0\rho}$ the normalized geomagnetic dipole moment ${\boldsymbol \mu}$ has the same dimension as velocity. Therefore,  the above used dimensional considerations Eqs. (12-13) can be applied for this case as well as their consequence Eq. (16). \\

  In paper Ref. \cite{cj} a power spectrum of the geomagnetic dipole moment for the period 0-1 Myr was computed using data from drift sediments in the Iceland Basin (Ocean Drilling Program - ODP, site 983 \cite{odp}). Figure 6 shows this spectrum in the log-log scales (the spectral data were taken from Fig. 6 of the Ref \cite{cj}).  The dashed curve in the Fig. 6 indicates the stretched exponential spectrum Eq. (16) corresponding to the helical distributed chaos (the analysis for a much longer-term period 0-160 Myr can be found in the Ref. \cite{b2}). \\

  The global magnetic solar activity dynamics can be described by the time series of the sunspot number, which is a scalar one. To understand the underlying magnetohydrodynamics, one needs a reconstruction of corresponding multi-dimensional phase space. It was estimated \cite{lamg} that for this purpose embedding dimension D=3 can be sufficient (see also Ref. \cite{data}). The solar magnetic field cycle is about 22 years (11 years magnetic field polarity reversals). This means that the underlying magnetohydrodynamics must have the corresponding symmetry group. Since the sunspot number time series does not possess such symmetry one should obtain a {\it cover} system (possessing the symmetry group) which is dynamically (locally) equivalent to the system without the symmetry group \cite{lg}. In the Ref. \cite{lamg} such cover system was constructed for the period 1750-2005 yy. Figure 7 shows the power spectrum for the reconstructed (cover) time series. The reconstructed data (cover time series) were taken from the site \cite{data} and the spectrum was computed using the Maximum Entropy Method.  The dashed curve in the Fig. 7 indicates the stretched exponential spectrum Eq. (16) corresponding to the helical distributed chaos.

\section{Acknowledgment}

I thank P. Odier for a consultation related to his paper.

\end{document}